# Wars Without Beginning or End:
## Violent Political Organizations and Irregular Warfare in the Sahel-Sahara

Olivier J. Walther
*University of Southern Denmark,* ow@sam.sdu.dk

Christian Leuprecht
*Royal Military College of Canada*

David Skillicorn
*Queen's University*

2 May 2016, 8800 words (+ bibliography and appendices)

**Abstract**
This article examines the structure and spatial patterns of violent political organizations in the Sahel-Sahara, a region characterized by growing political instability over the last 20 years. Drawing on a public collection of disaggregated data, the article uses network science to represent alliances and conflicts of 179 organizations that were involved in violent events between 1997 and 2014. To this end, we combine two spectral embedding techniques that have previously been considered separately: one for directed graphs (relationships are asymmetric), and one for signed graphs (relationships are positive or negative). Our result show that groups that are net attackers are indistinguishable at the level of their individual behavior, but clearly separate into pro- and anti-political violence based on the groups to which they are close. The second part of the article maps a series of 389 events related to nine Trans-Saharan Islamist groups between 2004 and 2014. Spatial analysis suggests that cross-border movement has intensified following the establishment of military bases by AQIM in Mali but reveals no evidence of a border 'sanctuary'. Owing to the transnational nature of conflict, the article shows that national management strategies and foreign military interventions have profoundly affected the movement of Islamist groups.

**Acknowledgments**
We are grateful to Larry Brooks, Dimitris Christopoulos, Alistair Edgar, Colin Flint, and Clionadh Raleigh. Alexandra Green and Marie Hugo-Persson provided valuable technical assistance. An earlier version of this paper was presented at the Balsillie School of International Affairs, Waterloo, on 19-20 March 2015, and at the Workshop on Subnational Governance and Conflict at the University of Sussex on 22 May 2015.

# 1. Introduction

Inter-state conflicts usually start with a declaration of war and end with surrender, negotiation or an armistice between the belligerents. By contrast, many modern conflicts are seemingly intractable. This trend is particularly evident in the Sahel-Sahara region in North and West Africa, a battleground for Islamists seeking to impose *sharia*, rebels seeking independence, transnational traffickers, and former colonial powers looking to project influence. While many Sahel-Saharan countries oscillate between periods of rebellion and political unrest, and periods of peace and reconciliation, the actual location and onset of violence defy prediction. Long considered a poster-child of political stability, Mali, for example, faced a military coup, a rebellion, a Western military intervention, and several major terrorist attacks – all in less than 2 years. By contrast, Chad, plagued by civil war and rebellion until the end of the 2000s, has been experiencing an unexpected period of stability – it is among the very few destinations in the Sahara that is still considered safe for tourists (OECD 2014).

Against this background, this article examines the relationships between alliances and conflicts as a putative explanation for the apparent unpredictability of many modern wars in the Sahel-Sahara. Social scientists have long been preoccupied with the logic of violence, explaining the presence and absence of violent conflict, and the onset and diffusion of internecine violence. Ostensibly, these are important questions to answer if we hope to prevent and forestall violent conflict. However, in the Sahara-Sahel these models seem to have little explanatory traction: although they vary, the preconditions for violent conflict are omnipresent, yet violence itself seemingly eludes prediction while actual patterns of violence are volatile. Recent research on the social determinants of conflicts in political geography has shown that political allegiances in the Sahel-Sahara were fluid and largely not ideologically motivated (Hagen 2014). Armed groups split and coalesce unpredictably, change names as new opportunities arise, and morph as ephemeral coalitions between tribal and ethnic groups. A similar volatility characterizes commanders and rank-and-file fighters who frequently shift allegiances among regular forces, rebel movements, and violent political organizations (VPOs), depending on local circumstances (Author, 2013).

The initial objective of this article posits a relational approach to the study of the structure of relationships among state and non-state actors. In doing so, it harnesses network theory for which social phenomena such as political violence are necessarily mediated by social interactions (Burt 1992). Drawing on a public collection of data on political violence, the article uses network science to represent alliances and conflicts between 179 organizations involved in violent events between 1997 and 2014. Owing to the fundamentally relational nature of internecine violence, we are particularly interested in the way the structural positions of conflicting parties affect their ability to resort to political violence. To this end, we combine two spectral embedding techniques that have previously been considered separately: one for directed graphs that takes into account the direction of relationships between belligerents, and one for signed graphs that takes into consideration whether relationships are positive or negative between groups.

Our second objective is to analyze the spatial patterns of violence across the region by localizing violent events and studying the geographic scales of regional dynamics. The literature presumes that borders matter because they circumscribe sovereign territories that impose transaction costs on those who cross. But do borders matter to Islamist groups in the Sahel-Sahara region? If so, what are their effects? And what transaction costs, countervailing or otherwise, do they impose on the movement of Islamist groups? In a geographical environment where populations remain poor and sparse, recent research on the spatiality of conflict in the region has established that cross-border mobility was an important component of military strategy (Zenn, 2014). Armed groups such as Al Qaeda in the Islamic Maghreb (AQIM) capitalize on marital, political, and financial ties throughout the region to attack targets, take hostages, and evade security forces (Guidère 2011, Wilkinson, 2012, Lebovich 2013, Bøås 2014). These general principles of desert insurgency based on mobility, speed and range challenge the Clausewitzian conception of warfare: hold territory and attack the enemy's strongest point (Keegan 1993). While the desert is seen by many as a hostile environment that "distances and isolates [VPOs] from major population centers and force them to disperse rather than concentrate their forces" (USAID 2014: 16), it can also provide a resource to strike anywhere, anytime, and without apparent logic. Focusing on 389 violent events in which nine Al-Qaeda-affiliated groups have been involved, we highlight specific spatial patterns that emerge from a longitudinal analysis of events over a ten-year period starting in 2004. Due to the transnational nature of conflict, we ascertain the countervailing transaction costs that borders represent, notably by testing whether national borders limit the displacement of Islamist groups or serve as sanctuaries whence attacks are launched.

The article proceeds as follows. The second section reviews the literature on the social and spatial organization of state and non-state organizations in West Africa, paying particular attention to the role of networks and national borders. The third section presents the data and explains how, using network analysis and geographical information systems, we structured them into networks and chronological events. The fourth section models the structural position of actors in conflict. The fifth section addresses the spatial patterns of Islamist groups and implications of the findings for theory, method, and practice.

## 2. Previous research

### 2.1 Conflicts and signed networks

Greater access to geo-referenced data and the use of spatial statistical analysis has advanced the study of social and spatial patterns of armed groups over the past decade. While past analyses of (civil) wars were limited by a lack of reliable data, the proliferation of satellite and disaggregated data has spawned innovative approaches to investigating the onset and diffusion of political violence across time and space (Zammit-Mangion et al. 2013, von Uexkull 2014, Dowd 2015, De Juan 2015). The concomitant proliferation of political and economic predictors on which the spatial-analytical approach in geography can draw now includes factors as diverse as the nature of government, ethnic divisions, poverty, income, inequality, number and morale of troops, frequency of droughts, and endowment of natural resources (O'Loughlin and Raleigh 2008 for a review). Some factors that may explain why groups resort to violence are also related to the structure of relationships that connect actors in conflict (Radil and Flint 2013, Philipps 2015). Political violence is a relational process; so, individuals and organizations are all, to varying degrees, embedded in networks of alliances and conflicts that enable or restrain action.

The increasing availability of disaggregated data, combined with recent conceptual and computational advances in network science, is reason to reevaluate the importance of social network analysis (SNA) within the field of conflict studies. SNA is the study of individual actors, groups, organizations or countries, represented by the nodes of the network, and the relationships between these actors, represented by their links. As both a paradigm of social interactions based on graph theory and a method, SNA seeks to understand networks by mapping out the ties between the various nodes as they are rather than how they ought to be or are expected to be (Newman 2010).

SNA is particularly adept at capturing the complexity of conflict situations due to its ability to describe, represent, and model signed networks, i.e. networks that contain both positive and negative relations. Positive ties develop to overcome collective-action problems, enforce trust and ideology, coordinate activities at a distance, distribute resources, or disseminate ideas and decisions. Political alliances between states are typical of positive-tie networks. By contrast, negative ties develop among actors that dislike, avoid, or fight one another with various levels of intensity. For positive and negative ties, SNA can be used to study the structure and function of the network as a whole, and the role of each node in the group in relation to others. Maoz and colleagues, for example, have used a network approach to model the effect of political and economic interdependence on the evolution of inter-state conflicts over a century (Maoz 2006, Maoz et al. 2006). Using the case of World War I, Flint et al. (2009) and Radil et al. (2013) have shown how alliances or rivalries between states could explain the diffusion of war on a global scale. Radil and Flint (2013) have applied the same approach at the African level.

Recent studies have noted that networks with positive ties tend to be structured differently from those with negative ties (Everett and Borgatti 2014). Networks based on friendship, alliance and collaboration are known for being denser and more clustered around actors that share similar values than networks containing negative ties, because individuals and organizations tend to have more friends than enemies (Huitsig et al. 2012). Positive-tie networks also convey more resources, ideas, and knowledge than negative-tie networks based on hatred, avoidance or conflict. As a result, many centrality measures based on the assumption that social networks serve as conduits for flows of information, advice, or influence, such as betweenness or closeness centrality, are unrealistic in the case of actors in conflict (Everett and Borgatti 2014). For example, if the Algerian military is the common enemy of both AQIM and the Movement for Oneness and Jihad in West Africa (MUJAO), flows are unlikely to pass from AQIM to MUJAO through the Algerian military. Networks containing negative ties are also well known for having a low level of transitivity, a principle that assumes that two actors that share a connection to a third actor, are likely to be connected themselves.

A growing literature in conflict studies and related social sciences suggests that, despite their differences, positive- and negative-tie networks should be analyzed simultaneously (Labianca and Brass 2006, Grosser et al. 2010, Rambaran et al. 2015, Yap and Harrigan 2015). One way to incorporate both allies and adversaries is to use structural balance theory, which argues that social relations are stable if they contain an even number of negative ties (Newman 2010). Stable groups of three actors (known as triads), for example, are theoretically stable if everyone likes everyone else, or if two actors are in conflict with a third party (Doreian and Krackhardt 2001, Hummon and Doreian 2003, Doreian and Mrvar 2015). Over time, unstable triads theoretically evolve towards stable triads, because instability creates tensions that can only be resolved by altering views, behaviors and alliances. Another approach to signed networks is to model the structural autonomy and constraints of actors. In a recent article, Smith et al. (2014) argue that an actor's political independence is constrained both by its potential to reach other actors' resources and by the structural position of allies and enemies. Being connected to a single ally that is free of threat considerably reduces the autonomy of actors in signed networks, while a diversified network of allies enhances autonomy.

This article adopts a complementary approach. Instead of assuming that political violence is explained by attributes of the belligerents or by exogenous factors, we posit network structure to enable or constrain political violence. To do so, the first part of our analysis aims at representing how VPOs are connected to their allies and enemies. Using several centrality measures, we identify subclusters of actors within which conflict or cooperation is particularly developed, and highlight the main structural differences between positive- and negative-tie networks. Since enemies and allies are inextricably linked in real-life networks, the second analytical part of the article considers positive and negative ties simultaneously. Using spectral embedding techniques that place the nodes representing organizations at the position that best balances the "pull" of allies against the "push" of enemies, we model the balance between the relative effects of having allies and foes simultaneously. We take into account the fundamental asymmetric nature of conflicts and consider whether groups attack more or less than they are attacked. Combining signed and directed networks, we expect groups with similar allies and foes and similar aggression patterns to form clusters that concur with their structural position in the social network.

### 2.2 Conflicts, borders and safe havens

A growing literature suggests that national borders are not only potential sources of dispute between states; they also provide a safe haven for VPOs that compete with sovereign states (for a recent review, see Medina and Hepner 2013). Borders open markets of opportunity for rebels and jihadists that exploit weak states (Korteweg and Ehrhardt 2006, Innes 2007, Campana and Ducol 2011) and their lack of intraregional cooperation. Of the 59 terrorist groups designated by the U.S. Department of State in January 2014, 39 are thought to operate from a safe haven (Arsenault and Bacon, 2015), many of them in border areas. In Asia, border sanctuaries are mainly found in Pakistan- and Indian-administered Kashmir, Pakistan's Federally Administered Tribal Areas, Pakistan's North Waziristan, and Baluchistan. In the Middle East, the Syrian and Iraqi borders are well-known safe havens to myriad armed groups (Philipps and Kamen 2014). Sunni-dominated areas of the Syrian-Lebanon border are used as safe havens by Syrian opposition forces, whereas Shia-dominated areas are used by Hezbollah to launch attacks or enter Syria (U.S. Department of State, 2014). In South America, the Colombia-Venezuela border and the Brazil-Paraguay-Argentina triangle have been linked to transnational groups (Brafman Kittner, 2007). In Africa, the Great Lakes region and the Liberia-Sierra Leone borderlands have been used by rebel groups to challenge the sovereignty of states, while Somalia is well-known as a sanctuary for Al-Shabaab (Meservey 2013, CSIS 2014).

How VPOs use borders in the Sahel-Sahara is a matter of scholarly debate. Some claim that terrorist groups such as AQIM have divided Africa into "zones", each led by an emir (Tawil 2010, Global Terror Watch 2013) and that the movements of the most prominent emirs are primarily centered on certain regions (GCTAT, 2013). This approach explicitly deems sovereign state borders safe havens for traffickers, rebels and terrorists. Others contend that the territorial notion of safe haven was probably an inappropriate concept to grasp the spatiality of Trans-Saharan politically violent organizations (Author 2014, 2015). While certain VPOs have repeatedly mounted attacks in the same area, this approach argues that attacks mostly seek to control strategic cities and lines of communication, and not to capture and hold territory. This principle explains why the aims of many Salafist groups are more socio-political than territorial: "All we want is the implementation of Sharia" reportedly said a member of the Ansar Dine group in Mali, "We are against independence" (The Punch 2012). This is reminiscent of pre-colonial polities that distinguished between ownership and control of land and where, as noted by Herbst (2002: 40), authority extended everywhere people had pledged obedience to the king. Daesh has adopted a similar strategy. It seeks to control a network of cities, roads, military bases and oil resources across Syria and Iraq and promotes the view that all true believers should be brought under a Caliphate, without necessarily holding a contiguous and fixed territorial entity (Arango and Barnardmay 2015).

This article engages this controversy in light of the spatiality of VPOs and the effect of sovereign state borders on their movements. Do borders represent sanctuaries behind which VPOs wage turf battles over aspirational homelands? Or, as T.E. Lawrence (1920) posited a century ago, is desert warfare more like naval war in the sense that insurgents are mobile and relatively indifferent to the constraints imposed by their environment? We propose that the degree to which groups can move and challenge the territorial integrity of sovereign states is a function of the porosity of borders. From the beginning of the millennium until the French-led Opération Serval regained control of the main cities of Mali in 2013, intraregional cooperation remained weak and each state had its own border-management strategy (Lacher 2012). This uncoordinated response to the rise of VPOs in West Africa should have an observable effect on their mobility patterns across the region. More cross-border movement might be anticipated where borders are not heavily guarded and/or easy to cross due to informal arrangements with state authorities than where there is greater capacity to ensure border integrity through effective border patrols and law enforcement.

## 3. Research design

### 3.1. Social network analysis

Our analysis relies on disaggregated data from the Armed Conflict Location and Event Dataset (ACLED), which provides a comprehensive list of political events by country between 1997 and 2014 (ACLED 2015, Raleigh et al. 2010, Raleigh and Dowd 2015). The fifth version of the data was used to select 37 VPOs in the Sahel-Sahara region, their allies and their enemies, excluding non-identified Islamist and Libyan militias (see Appendix 1). The scope was limited to events with the following seven referents: Battle – no change of territory; Battle – Non-state actor overtakes territory; Battle – Government regains territory; Riots and protests; Violence against civilians; and Remote violence. This produced a list of 3231 events involving 179 organizations and 27,791 fatalities.

The ACLED dataset describes (up to) four groups in each incident: an attacker (A), a collaborator in the attack (B), a target (C), and a potentially assisting group that may also be a secondary target (D). This data is used to build a social network in which the nodes are groups, with positively weighted directed ties between groups that interacted (B to A, and D to C) and negatively weighted directed ties between attacker and target (B to C). For example, on January 12, 2014, clashes between French troops (A) and Malian troops (B) on the one hand, and Ansar Dine (C) and MUJAO (D) on the other hand, claimed 11 lives, including Islamist leader Abdel Krim, and 60 injured (ACLED incident 486MLI). Incidents are aggregated so that the ties between any pair of groups reflect all of their interactions. There can be both positive and negative ties, and in both directions, between the same two groups. Directions are important because they are surrogates for intentionality: a group on the offensive makes a conscious decision to attack while the defender has no choice, and other groups must decide whether to join in or not. These decisions reflect a calculus of advantage or ideological alignment.

The resulting graph is analyzed in two steps. First, we map the networks containing negative and positive ties separately and analyze the most prominent actors using several centrality measures. Because negative-tie networks do not serve as conduits for flows of information, advice, or influence, we use degree centrality, which simply refers to the standardized number of ties each node has, and eigenvector centrality, which refers to the number of nodes adjacent to a given node, weighted by centrality, and indicate whether nodes are connected to other well-connected nodes. For our positive-tie network, we use eigenvector centrality and betweenness centrality, which measures the number of shortest paths from all nodes to all others that pass through that node (Freeman 1979).

Second, we combine both positive and negative ties into a single network, and embed this network in a geometric space in such a way that the distance between each pair of points accurately reflects the balance between the 'pull' from collaborating groups and the 'push' from aggression between them. These distances are globally integrated by considering not only immediate neighbors[1], but neighbors of neighbors and, in fact, the structure of the entire graph. It is this integration that makes the process challenging: positive relationships are naturally transitive ("the ally of my ally could plausibly become my ally") but negative relationships are not (the proverbial "enemy of my enemy is my friend" does not obtain). Technically, the adjacency matrices that describe positive and negative ties are combined into a matrix, called a Laplacian, that combines both kinds of ties and normalizes the representation so that well-connected nodes are central and poorly connected nodes peripheral (see representation Zheng and Skillicorn 2015, Zheng et al. 2015 for more details). This Laplacian matrix is transformed to discover the directions in which the graph varies the most and these are used as axes for creating an embedded graph. In this representation, position and distance are meaningful. Sets of 'bad actors' such as VPOs and (supposedly) 'good actors' such as governmental forces and civil society tend to form polar opposites in some dimension(s) of the representation. Since proximity represents similarity – commonly known as an alliance – distance tends to represent opposition.

---

[1] In the networks considered, "neighbors" refer to actors who cooperate or fight each other, and do not necessarily refer to locational space.

### 3.2. Spatial patterns

To understand the spatial strategies of VPOs, we limit the scope to nine Trans-Saharan organizations that subscribe to an Islamist ideology: Al Qaeda, Ansar Dine, AQIM, the Armed Islamic Group (GIA), Al Mourabitoune, the Free Salafist Group (GSL), the Salafist Group for Preaching and Combat (GSPC), MUJAO, and Those Who Signed in Blood. Affiliated with al Qaeda, these Islamist groups share a common historical and ideological background and form several components of a single, flexible network, rather than independent entities. As reminded by Hagen (2014: 2), "AQIM overlaps with a number of nominally independent and 'locally-focused' groups, such as Ansar al Dine and MUJAO. These groups are part of the larger AQ family and cannot be separated from AQ and AQIM". Mergers, name changes and splits are common. For example, GSPC – a splinter group of GIA of Algeria – rebranded itself as AQIM in 2007. Some of its members broke off in 2011 to form MUJAO while others formed Al Moulathamoun (2012) and Al Mouakaoune Biddam (2012) (Wojtanik 2015). In 2013, MUJAO merged with Al Moulathamoun to form Al Mourabitoune, which, in 2015, was renamed Al Qaeda in West Africa (Joscelyn 2015). Those groups frequently exchange information, funding, hostages, and conduct joint operations since several of their leaders have historically been members of AQIM's leadership network (Hagen 2014) and have developed multiple allegiances across organizations (United Nations 2015). Our sample does not include Boko Haram; its activities are almost exclusively located in the Sahelian part of West Africa and its use of political violence differs from other Islamist organizations in the region (Menner 2014, Zenn 2015). While AQIM has urged local emirs to refrain from violence against civilians and encouraged them to gain the hearts of the people, Boko Haram has, since its leader was killed in 2009, used indiscriminate violence as its principal political instrument (Forest 2012, Zenn 2012).

Using a Geographical Information System (GIS), our first step is to map all violent events in which these Islamist groups were involved over the last 10 years. The year 2004 is the starting point of the development of VPOs that had hitherto been almost exclusively located in Algeria (Dowd and Raleigh 2013). These groups, their allies and their enemies are responsible for 389 violent events totaling 1,434 fatalities through 2014. Our next step is to develop two hypothetical scenarios to distinguish distinct spatial patterns. The 'mobility' scenario assumes violence follows a linear chain of events: groups move from one location to the next, possibly across borders, without returning to their original location. This would reflect the strategy of a group under intense pressure from security forces, or, alternatively, of a group that has mastered movement in an arid environment. By contrast, the 'sanctuary' scenario supposes a territorial turf whence groups operate across borders and a clear origin of flows. Once the location of each event is known, our third step is to connect violent events chronologically through hypothetical lines and verify if the general spatial pattern of the attacks corresponds to one of the two scenarios described above: the 'mobility' scenario where groups move freely across borders, or the 'sanctuary' scenario where groups use a particular region as a rear base. Since the ACLED database does not contains information about the movements of Islamist groups, we use dotted lines to indicate that spatial patterns based on the location of violent events do not necessarily correspond to actual physical movement between places, but rather to a longitudinal series of events.

## 4. A social network analysis of political violence

### 4.1. Negative- and positive-tie networks

We start with a graph that represents each organization as a node that is connected to those actors with which it is in conflict. The size of the nodes in Figure 1 is proportional to the number of ties (or degree).

Figure 1: Negative ties between organizations involved in violent events, 1997-2014

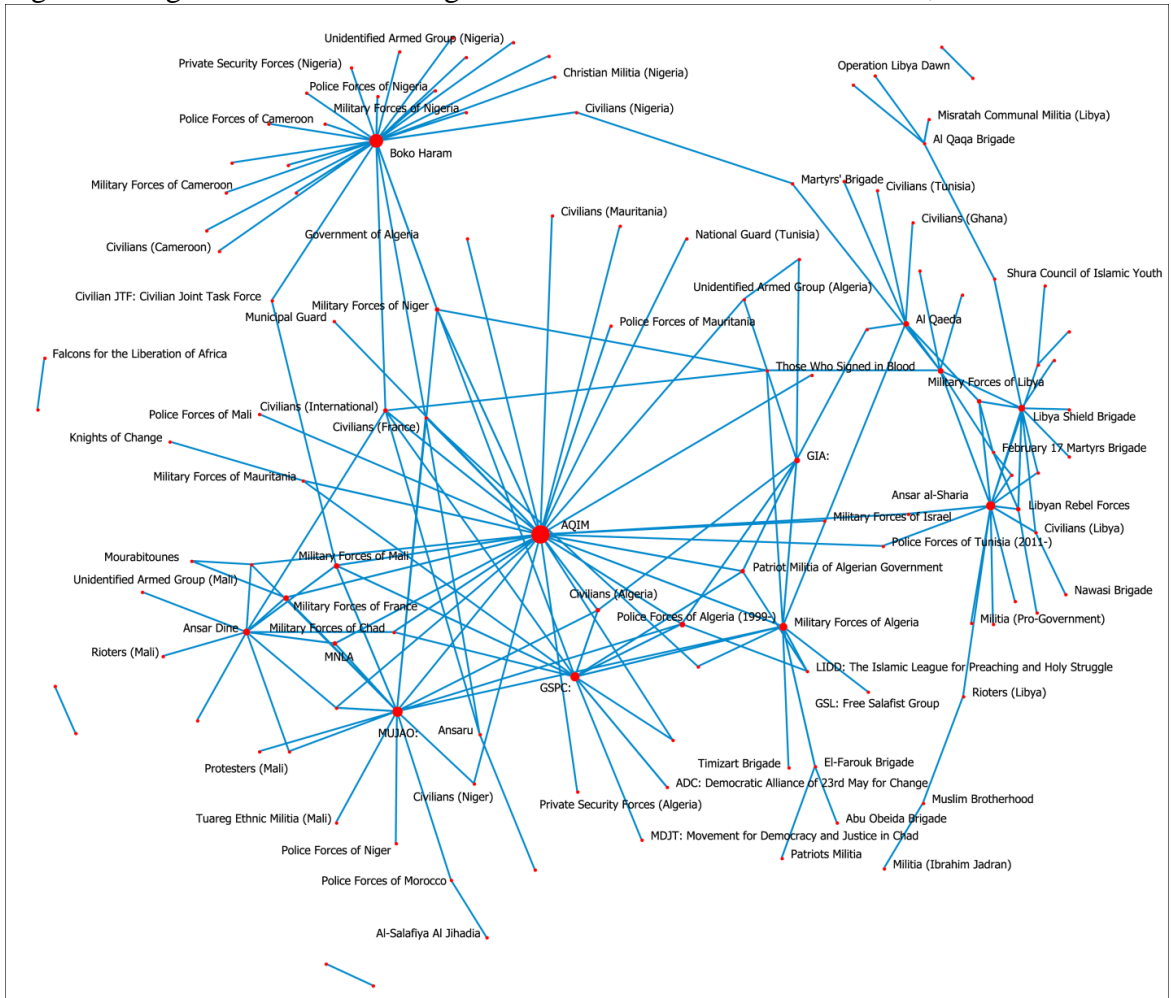

Note: Isolates are not shown.

Three main clusters emerge: the Nigerian cluster that is polarized by Boko Haram; the Trans-Saharan cluster that is composed of groups affiliated with Al Qaeda such as GSPC and AQIM and their enemies; and the Libyan cluster that is composed of myriad Islamist brigades and pro-government forces. With a density of only 0.023, the network is very sparse, which is typical of networks that are made up exclusively of negative ties: the number of enemies a group can have is often more limited than the number of potential allies (Huitsig et al. 2012). The network also has a low level of transitivity: in only 1.2% of the triads enemies of enemies are in fact enemies, while in most cases (98.8%), enemies of enemies are friends. Finally, organizations with adverse attributes tend to be in conflict with one other, a tendency known as heterophily (Csaba and Pál 2010, van Mastrigt and Carrington 2014). This can be tested using the E/I index, which calculates the difference between external and internal ties for each group of actors (government, rebels, militias, civilians, Islamists, and external forces), divided by the total number of ties. The E/I index for the network is positive (0.899) and statistically significant (chances of getting the result right by guessing are less than 1%), which confirms that VPOs clash with organizations that do not belong to the same category.

At the level of the organization, the network is composed of few highly central organizations (Table 1). That makes sense since being in conflict with many adversaries simultaneously is widely regarded as a liability rather than as an asset (Labianca and Brass 2006). Negative relationships adversely affect the outcomes of VPOs' military operations, reduce their ability to coordinate activities across the region, and limit their ability to cooperate to achieve their political or religious goals. Among politically violent organizations, AQIM has the highest score in degree and eigenvector centrality, which indicates that it has the greatest number of enemies and is connected to other actors that also have many enemies, such as the military and police forces of Algeria. MUJAO, GSPC and GIA also occupy a prominent structural position due to their conflicts with civilians, and armed forces in several countries. Other prominent actors include Boko Haram, which stands out for being connected to many other actors who themselves have few connections to one other, and some Libyan groups such as Ansar al-Sharia and Libya Shield Brigade.

Table 1: Top-scoring nodes for selected centrality measures – negative ties

| Rank | Degree centrality | Eigenvector centrality |
|------|-------------------|------------------------|
| 1 | AQIM (0.264) | AQIM (0.743) |
| 2 | Boko Haram (0.200) | MUJAO (0.421) |
| 3 | MUJAO (0.136) | Military Forces of Algeria (0.289) |
| 4 | Ansar al-Sharia (0.120) | GSPC (0.257) |
| 5 | Ansar Dine (0.096) | Ansar Dine (0.229) |
| Mean | 0.024 | 0.071 |
| Std. Dev. | 0.035 | 0.095 |

Note: Scores are indicated between brackets.

The structure of the network of enemies contrasts strongly with the one showing how organizations involved in violent events have collaborated across the region. As depicted in Figure 2, the positive-tie network is divided into three main unconnected groups of allies, one triad connecting an unidentified armed group to Boko Haram and Ansaru, and three dyads.

The main cluster on the left is structured around North and West African military and police forces and their civilian allies, which are represented in red and yellow respectively. This cluster is indirectly connected to some of the main Islamist groups in the region, which are represented in green, through the secessionist movement MNLA. MNLA was allied with Ansar Dine in the first weeks of the Malian conflict before switching sides and fighting alongside the French-led military forces in 2013. The two other clusters are related to the Libyan conflict. One is structured around the armed forces of Libya and their pro-government brigades and battalions, the other around Islamist groups and ethnic and communal militias. Each cluster has a chain-like structure in which organizations are rather distant from one another. The Algerian Private Security Forces, for example, are eight steps away from Al Mourabitoune. The long path-length distance, low density (0.034) and low clustering coefficient (0.104) of the network are typical of a structure that is not organized around groups of tightly connected actors. This suggests that most governmental forces and VPOs tend to build bilateral or trilateral alliances rather broad coalitions across the region. The graph also highlights the lack of regional cooperation between government forces that face similar threats: there is no reported tie between the military forces of Libya and Algeria, or between the military forces of Cameroon and Nigeria.

Figure 2: Positive ties between organizations involved in violent events, 1997-2014

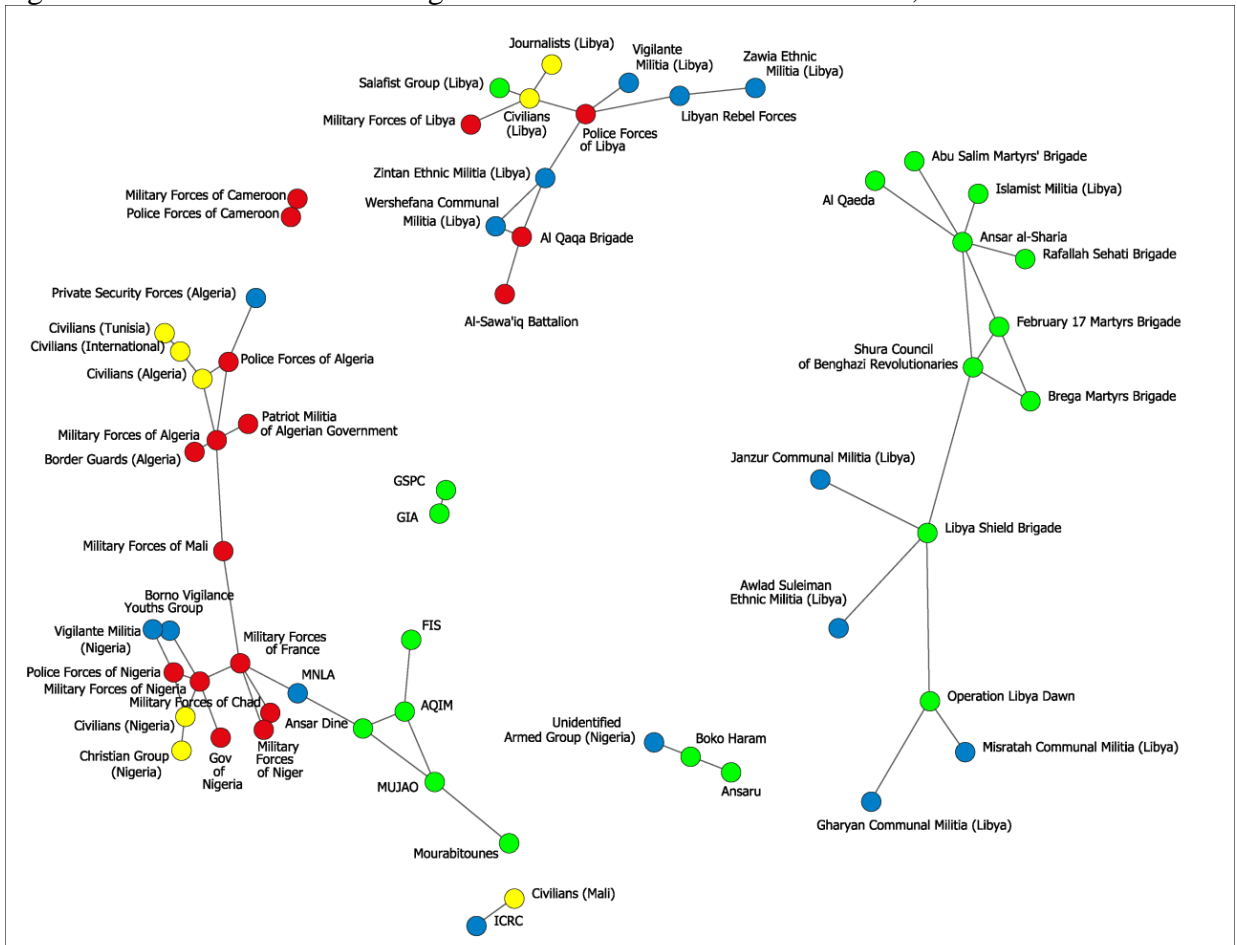

Notes: green nodes refer to Islamist groups, red to government forces, yellow to civilians, and blue to other actors.

Military and police forces have the highest eigenvector and betweenness centrality, followed by Ansar al-Sharia and the Shura Council of Benghazi Revolutionaries (BSCR), both of which hail from Libya (Table 2). Generally speaking, betweenness centrality scores – that refer to the propensity to bridge clusters – are very low, even for top-scoring nodes, which suggests that the networks contain few exceptional brokers. Only the French military forces play a role in bridging several African armed forces that would otherwise not be connected, hence their high betweenness centrality. Once again, the isolation of Boko Haram in Nigeria contrasts sharply with the network of alliances among other Sahelo-Saharan and Libyan groups.

Table 2: Top-scoring nodes for selected centrality measures – positive ties

| Rank | Eigenvector centrality | Betweenness centrality |
|---|---|---|
| 1 | Military Forces of Nigeria (0.400) | Military Forces of France (0.111) |
| 2 | Police Forces of Nigeria (0.379) | Military Forces of Algeria (0.071) |
| 3 | Ansar al-Sharia (0.223) | Military Forces of Mali (0.070) |
| 4 | Shura Council of Benghazi Revolutionaries (0.260) | Military Forces of Nigeria (0.062) |
| 5 | Military Forces of Libya (0.193) | MNLA (0.052) |
| Mean | 0.039 | 0.012 |
| St. Dev. | 0.082 | 0.022 |

Note: Scores are indicated between brackets.

### 4.2. Spectral embedding

Now, we compute the spectral embedding of the social networks derived from the ACLED data. Initially, for the sake of simplicity, we disregard the direction of the ties. The embeddings are shown in Figure 3. Negative ties resulting from recorded attacks are shown in red and positive ties resulting from alliances, or at least common purpose, are shown in green. The general structure is of a group of opposing poles representing groups whose primary relationship is that they attack or are attacked by groups at the other pole. The graph clearly shows how 'bad' actors such as Islamist and Jihadist groups are grouped opposite 'good' actors, both violent and non-violent. The contrast is particularly evident for Boko Haram, and its opposition to governmental forces and civilians from Nigeria and Cameroon, as well as for GIA-GSPC-AQIM, and its opposition to Algerian armed forces and civilians. The graph also shows that the attack patterns of GIA, GSPC and AQMI differ significantly from those of Ansar Dine, MUJAO and Al Mourabitoune, which are located much closer to the center of Figure 3.

Figure 3: Spectral embedding showing positive and negative ties

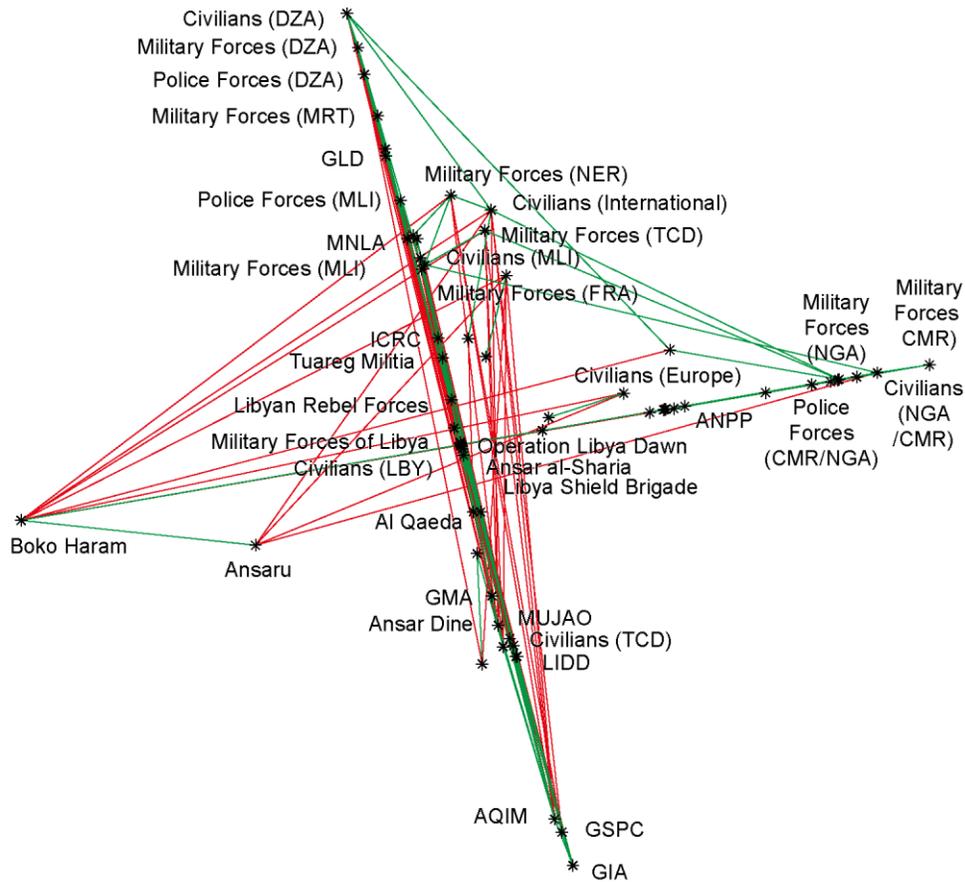

Any measure that considers a group in isolation is unable to distinguish VPOs from military or police organizations because both have similar patterns of interaction. We, therefore, compute measures of outward and inward aggression based not on the number of such incidents but on the length of the relevant ties in the embeddings. A group's position in the embedding reflects its relationships with all of the groups with which it interacts, and, therefore, the length of the embedded ties is more revealing than simply the number of attacks. For example, the distance of a group from the center of the embedding reflects not only how many other groups attack it (or are attacked by it) but also the extent to which its enemies are similar to one another (close in the embedding). Thus a long red tie measures not only the existence and frequency of attacks, but also their strategic intensity. On Figure 4, groups are plotted at the same positions as in the spectral embedding presented in Figure 3 to denote "levels of aggression". They measure the outgoing aggression of each group and the incoming aggression to which it is subjected. They are color-coded: red means a group generates more aggression than it receives; orange means that the group generates some outgoing aggression; and green means that there is no outgoing aggression (individual scores are presented in Appendix 1).

Figure 4: Spectral embedding showing levels of aggression

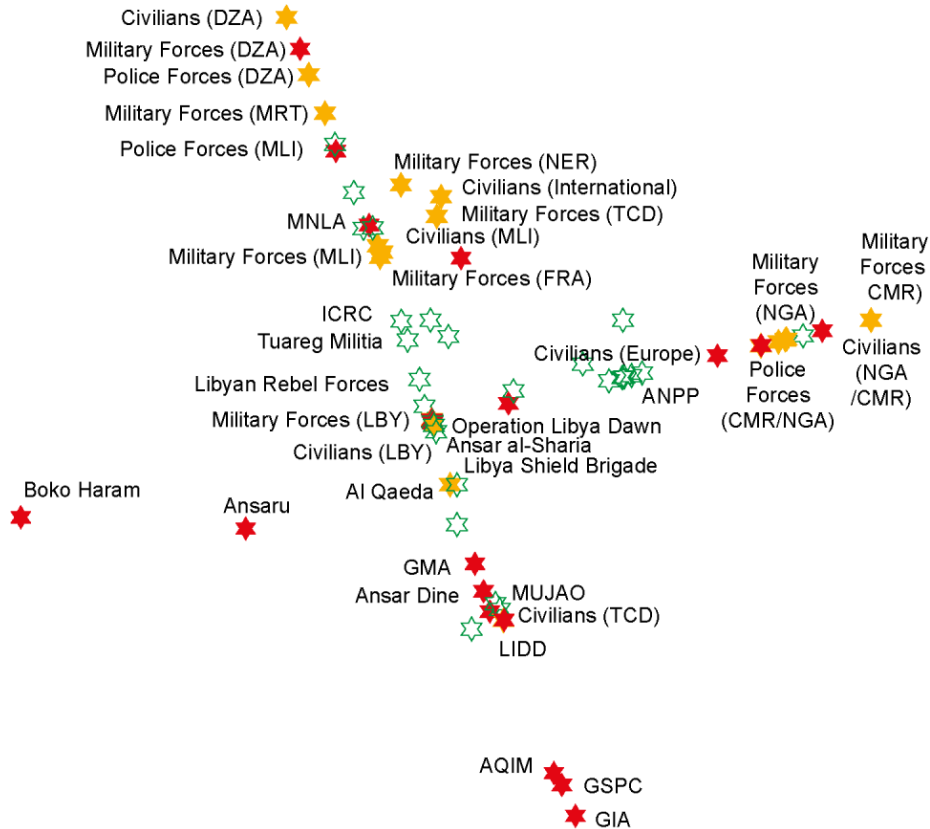

The vicinity of groups presented in Figure 4 allows us to distinguish VPOs (almost all groups are red) from national defense forces (many are also red but adjoined by orange or green). In other words, most of the polar opposites consist of 'bad actors' on one side and 'good actors' on the other, and they are structurally distinct. 'Bad' aggressive actors such as AQIM or Boko Haram tend to be in clusters of net aggressors, or isolated; 'good' aggressive actors such as militaries tend to be in clusters with orange and green groups. Neutral actors such as the International Committee of the Red Cross (ICRC) tend to fall in the middle and green. Victims are also green but tend to be located near their champions.

Northern Nigeria and Libya are particularly interesting as they involve many VPOs with strong structural constraints. From the literature we would expect Northern Nigeria, where Boko Haram is particularly dominant, to have more of a dual structure than Libya, where myriad of violent groups compete for the control of the state and oil resources (Gow et al. 2013). We find that our intuition was correct as Figures 5 and 6 show. Spectral embedding showing conflicts and cooperation for 37 organizations in Northern Nigeria clearly confirms that Boko Haram is in conflict with virtually everyone, a situation comparable to that of ISIS in the Middle East, which opposes all governments and non-state actors – including Al Qaeda – in the region.

Figure 5: Spectral embedding showing positive and negative ties for 37 organization in Northern Nigeria

Figure 6: Spectral embedding showing positive and negative ties for 30 organizations in Libya

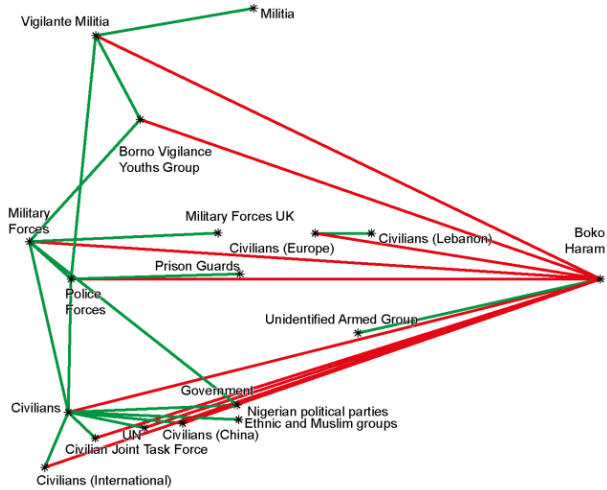
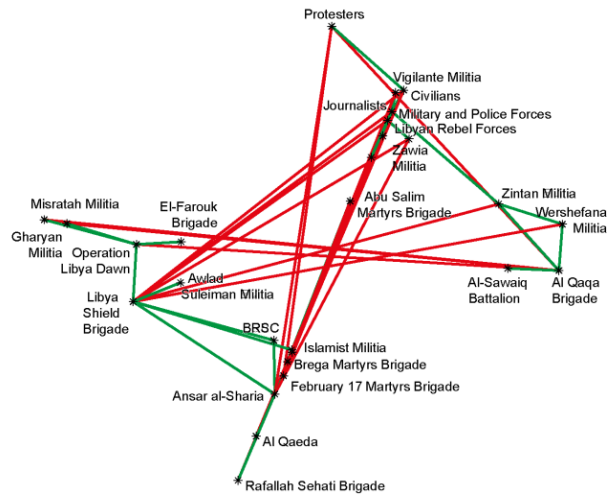

Note: for the sake of clarity, Ansaru is not shown.

In Libya, spectral embedding conducted on 30 organizations highlights the ongoing conflict between pro-Islamist groups and pro-government forces (Figure 6). Islamist groups, on the left of the graph, are composed of Islamist militias such as Libya Dawn and Libya Shield, and of Jihadist groups close to Al Qaeda such as the Revolutionaries Shura Council (BRSC), a coalition that includes Ansar al-Sharia, the 17 February Brigade, and the Rafallah Sehati Brigade. These groups, based in Tripoli and in Benghazi, all oppose the Libyan army, as indicated by several long red ties. Among pro-government forces, on the right, are anti-Islamist militias such as the Zintan Militia, the Al-Sawaiq Battalion and the Al Qaqa Brigade. Civilians and journalists are located near the internationally recognized authorities of Libya.

## 5. Spatial analysis of mobility patterns

This section examines the spatiality of select Islamist groups that have developed attack patterns across the Sahel-Sahara region. For strategic and policy purposes, we are particularly interested in whether state border strategies and multinational military missions have had a measurable effect on the transborder movement of Islamist groups.

### 5.1. Changing mobility patterns

The analysis that follows reveals no evidence of a 'sanctuary' pattern in which Islamist groups make systematic use of a particular border area. However, Table 3 reveals that the movement of Islamist groups changed completely between 2004 and 2014: while the first seven years were marked by an apparent unpredictability of events across time and space, the last three years were characterized by a concentration of events, due to the outbreak of the Malian conflict and the strategies adopted by some states to control their borders. This shift occurred in a political environment where the number of violent incidents and victims related to Islamist groups in Sub-Saharan Africa has been on the rise (Dowd 2013, 2015, Dowd and Raleigh 2013). While only 30 events totaling 201 victims were reported between 2004 and 2008, there were 359 events and 1233 victims between 2009 and 2014. The intensification of attacks significantly reduced the average frequency between violent incidents, from an average of 44.8 days in 2008 to 7.5 days in 2014.

Table 3: Key metrics

| Year | Number of events | Cross-border movements (%) | Number of victims | Average distance between events (km) | Average distance to borders (km) | Average time between events (days) |
|------|------|------|------|------|------|------|
| 2004 | 15 | 43 | 103 | 478 | 132 | 27.3 |
| 2005 | 3 | 50 | 63 | 502 | 39 | 122.7 |
| 2006 | 3 | 50 | 15 | 675 | 334 | 209.0 |
| 2007 | 1 | - | 3 | - | 162 | 35.0 |
| 2008 | 8 | 43 | 17 | 411 | 63 | 44.8 |
| 2009 | 11 | 80 | 30 | 136 | 137 | 36.5 |
| 2010 | 17 | 63 | 75 | 708 | 199 | 18.4 |
| 2011 | 29 | 61 | 64 | 864 | 106 | 12.8 |
| 2012 | 103 | 7 | 153 | 234 | 148 | 3.5 |
| 2013 | 155 | 13 | 768 | 330 | 146 | 2.6 |
| 2014 | 44 | 19 | 143 | 463 | 154 | 7.5 |

The region has always been characterized by a high level of transborder activity. Figure 7 confirms that, until 2011, Islamist groups travelled extensively across borders and, in many regions of Mali, Mauritania, Algeria and Niger, without much risk of being apprehended. After Algeria expelled them, they were tolerated by the Malian government of President Amadou Toumani Touré (2002-2012), which sought to capitalize on divisions within Tuareg society and on a withdrawal of the state to administer the northern part of the country. Successive events repeatedly occurred hundreds or thousands of kilometers apart, in different countries, and irregularly, from Algeria to Mauritania, the Mauritanian-Malian border, and Niger. In 2005 and 2006, the average distance between two events exceeded 500 km, which is impressive given that harsh terrain and lack of road infrastructure, in particular in the Sahara.

One of the best known movements of this period is also the one that marked the beginning of the Saharan expansion of what would become AQIM. Between 21 February and 11 April 2003, 32 European tourists were kidnapped in the region between Illizi and Amguid in Algeria by Abderazak el-Para (born Amar Saïfi) and Abdelhamid Abu Zeid (born Mohamed Ghadir), two militants of GSPC. As Algerian security forces gave chase, the terrorists and hostages initially journeyed of over 3000 km to northern Mali. After having spent several months establishing alliances with leaders of local nomadic tribes, they moved to Niger through the plains of Azawagh, Aïr Mountains and the Ténéré desert, and ended up in the mountainous area of Tibesti in Chad where they were killed or captured, a second journey of over 2500 km through some of the most inhospitable environment on the planet (Author 2010).

In 2011, Mauritania and Algeria undertook a series of joint counter-terrorism operations aimed at AQIM's military bases. Such an attack took place in the Wagadu forest on the border between Mauritania and Mali in June. The central intelligence cell created to facilitate co-ordination between Saharan and Sahelian countries, known as the Combined Operational General Staff Committee (CEMOC), first met in Bamako in April 2011. Nonetheless, the level of regional cooperation remained low because Mali was not trusted by its neighbors, which accused it of colluding with Islamist groups. Henceforth, Mauritania and Algeria would conduct military operations in Mali when they deemed their interests to be threatened by the activities of transnational groups. The chronological succession of attacks by AQIM in 2011 shows a high intensity and percentage of cross-border movements. For example, AQIM claimed responsibility for a bomb attack in Bamako, the capital of Mali, on 5 January, followed by a hostage taking in Niamey, Niger three days later. On February 1, these attacks were followed by an AQIM car bomb in the Mauritanian town of Adel Bagrou, the abduction of an Italian tourist in Djanet, Algeria, a day later, and the killing of a Mauritanian policeman by two members of AQIM in the region of Legsseiba near the north bank of the River Senegal on February 3.

The year 2012 contrasts sharply with the period 2004-2011 because most events transpired in Mali, and, to a lesser extent, Algeria. Following the fall of Col. Muammar Gaddafi in Libya (2011) and President Amadou Toumani Touré (2012) a provisional alliance between Al Qaeda-affiliated groups and secessionists rebels of the MNLA launched a wide-ranging military offensive against the Malian army. Over a matter of weeks, all major cities of Northern Mali were seized, including Tessalit and Kidal in the Adrar des Ifoghas, where the offensive started, as well as Menaka, Timbuktu and Gao. New groups such as MUJOA and Ansar Dine were particularly active during this period and started to clash with their former Tuareg allies over the cities of the north of the country and main lines of communication. Our analysis shows that during this period the distance between violent events and borders is the more stable (approximately 150 km) than the preceding period during which average distances to borders varied from 39 to 334 km.

In 2013, the French-led Opération Serval reasserted control over Northern Mali. As French and Chadian troops progressed north, Islamist groups were driven from Kona, Douentza, Gao, Timbuktu, and were chased out of their stronghold of the Adrar of the Ifoghas. Operation Panthère, launched around Tessalit on 18 February 2013, successfully defeated them, possibly because the French and their allies adopted some of the principles of warfare that had made Islamists and rebels so successful in the region. Operation Panthère relied on a combination of airstrikes, artillery and ground combat operations conducted by the French, knowledge of the country provided by Tuareg guides, and Chadian desert warfare. Chadian troops and their highly mobile light trucks proved as effective in Mali as they have in the past 30 years in their own country.

Figure 7: Events connected chronologically, 2004-2014

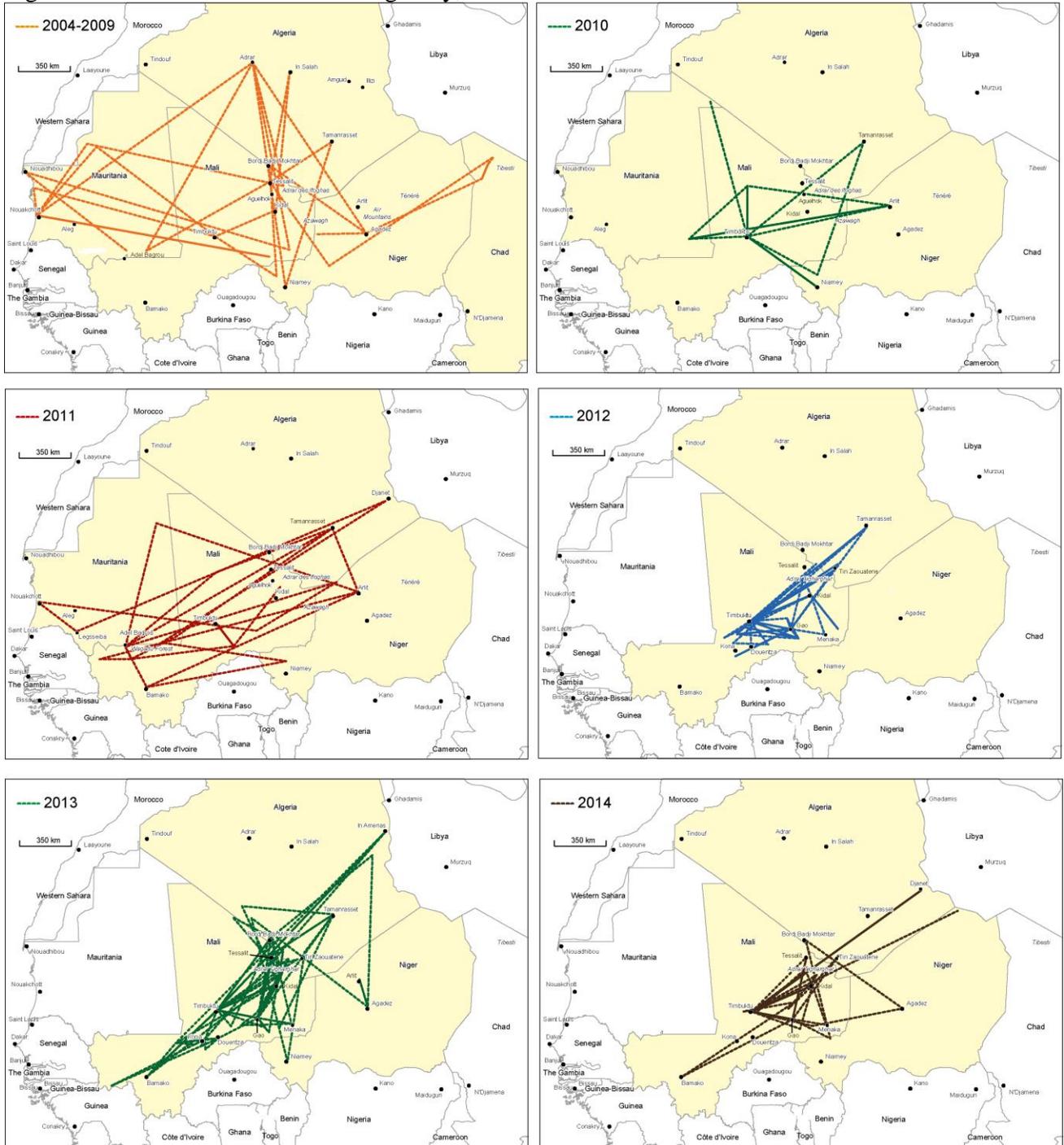

Note: the color of the lines refers to the year(s).

As in 2012, most events in 2013 took place in Mali and Algeria, along a south-west–north-east axis extending from Bamako in Mali to Tamanrasset in Algeria. While the French and their allies pushed north, the rebels of the MNLA seized Kidal and Tin Zaouaten from Islamist groups, and clashed with AQIM and MUJAO. The Algerian army also clashed with Islamist groups fleeing Mali towards Libya. The most brazen terrorist attack was launched in January 2013 against the gas facility of In Amenas in Algeria where MUJAO and Al Moulathamoun coordinated their activities, resulting in at least 67 deaths. A decade after being expelled from Algeria, Al Qaeda-affiliated groups were back in the country. Later that year, a camp of Niger's military was hit by MUJAO in Agades in May, military barracks were attacked in Niamey in June by Those Who Sign in Blood, and two French journalists were killed by AQIM in Bamako in November 2013.

The spatial patterns of attacks in 2014 are similar: principally concentrated in Northern Mali and, to a lesser extent, in Southern Algeria, localized events resulted from the French offensive in the Tighaghar Mountains that killed many Islamist leaders, and from MNLA rebels clashing with MUJAO. Roadside bombs and suicide car bombings organized by Al Mourabitoune targeted the United Nations Multidimensional Integrated Stabilization Mission in Mali (MINUSMA). The French-led anti-terrorist Opération Barkhane replaced Opération Serval in August 2014, and relies on highly mobile forces. Contrary to previous military engagements that targeted "one country, one crisis, and one theatre of operations", Opération Barkhane explicitly addresses the regional and cross-border dimension of terrorist activity throughout the region. The Operation relies on three ports in the Gulf of Guinea, two main airports in the Sahel, and a series of Saharan outposts located at the extreme periphery of Chad, Mali and Niger to disrupt cross-border trafficking routes and terrorist networks. Considering that Algeria and Libya are beyond the reach of any foreign armed forces and that the French military is stretched to the limit (de Galbert 2015), for the time being Opération Barkhane is the most ambitious military initiative at the regional level.

### 5.2. State border strategies

Do borders not matter if they cannot offer sanctuary? On the contrary, our analysis suggests that borders do matter because countries have put in place management strategies that affect the spatial patterns of Islamist groups. Building on Arsenault and Bacon's (2015) distinction between government will and capability to fight foreign terrorist groups, four situations can be distinguished (Table 4). States may have the military capacity to challenge transnational groups and choose to do so. In the region, Mauritania is probably the country most similar to this situation. Despite the small size of its armed forces, Mauritania took a number of strong military measures to control the flows of Islamist groups across its borders: develop mobile patrols, attack intruders systematically and establish partnerships with local tribes. As our analysis revealed, this firm attitude, combined with a national strategy for deradicalization (Ould Ahmed Salem 2013), has occasioned a significant decrease of violent events since 2012.

Table 4: Government will and capacity to fight transnational groups

|  |  | Will | |
| --- | --- | --- | --- |
|  |  | Strong | Weak |
| **Capacity** | Strong | e.g. Mauritania | Algeria |
|  | Weak | e.g. Niger | e.g. Mali |

Some states may also be in a position to deny foreign fighters access to their territory but choose not to do so, for strategic reasons. Whether Algeria would have adopted this strategy is a matter of debate. On the one hand, having suffered from decades of political violence, Algeria is strongly committed to combating terrorism in the region and plays a major role in peace negotiations in Mali. During the Malian conflict, Algeria deployed tens of thousands of troops to secure its borders, which are probably the most heavily guarded in the region. On the other hand, Islamist groups would have been unlikely to emerge in Northern Mali if Algeria's borders had been hermetically sealed. Many of the aforementioned groups originally hail from Algeria or are affiliated with organizations based there, and there is strong evidence that Algerian borders are easily crossed by Islamist groups to get food and oil supplies through informal arrangements with state representatives (CSIS 2014).

The third situation, where states may want to disrupt transnational groups without being able to project military power into borderlands, characterizes Niger. The 5,700 km border of Niger has long been poorly guarded due to lack of men and material. AQIM and other groups have capitalized on the situation either to take hostages in the region or to move to the south of Libya, where numerous Islamist groups have found favorable conditions (CSIS 2014). Finally, Mali exemplifies a lack of political will and ability to fight foreign Islamist groups. The gradual withdrawal of the state from the North of the country in the 2000s left Mali's borders unguarded and frequented by traffickers and militant groups.

## 6. Conclusion

The article examines the structure and spatial patterns of VPOs in the Sahel-Sahara, a region characterized by growing political instability over the last 20 years. Based on a novel approach that combines signed and directed graphs, our methodological contribution has been to highlight opposed groups and distinguish among several kinds of aggressors depending on their conflict patterns. In settings where groups form shifting alliances and oppositions, an approach that takes into account not only the local, pairwise relationships, but also the global patterns that emerge, is needed for situational awareness. Conventional social network analysis can represent positive ties, but not ties where direction matters, or where ties represent a negative association. Furthermore, these are not independent properties of a social network and so must be represented together. This article has illustrated the effectiveness of extending social network analysis to such settings. Conventional social network measures fail in these settings. For example, measures such as betweenness are inappropriate because negativity does 'flow' in the way that positivity is conceived, and centrality is not a crucial property when negativity separates nodes far from the center.

From a theoretical perspective, the paper advances theory on the spatiality of Islamist groups by showing that sanctuary is immaterial to Saharan borders because many Islamist groups seek to control the movement of people and (often illicit) goods (Bøås 2015). The inability to garrison a sparsely populated region such as the Sahel-Sahara – which is the size of the United States – makes it difficult to hold territory. This situation is similar to the one adopted by Daesh between Syria and Iraq, and radically different from the territorial objective of such groups as Boko Haram in Africa, or the Taliban on the Afghanistan-Pakistan border, for which the defense of a delineated territory is paramount.

Building on publicly available data, we started by mapping how 179 organizations involved in political violence were structurally connected through conflict and alliances. Our results show that the network that connects actors in conflict has a low density, a low level of transitivity, and contains few central actors, three typical features of negative-tie networks. AQIM is unequivocally the most connected organization, both in terms of the overall number of actors with which the group is in conflict, and the respective centrality of its enemies. In network terms, this is a liability. Divided into several clusters, the positive-tie network has a long path-length distance, low density and low clustering coefficient, a structure that suggests that most organizations tend to build limited alliances rather than broad coalitions across the region.

We then combined the two networks and modeled the effect of having friends and foes simultaneously. Using the attack relationships, we also measured the level of outgoing and incoming aggression of each group. From this approach, five categories emerge. The first category includes neutral actors, represented in the middle of our graphs. The following categories include three kinds of groups that cluster together: victims, groups that are attacked more than they themselves attack, and groups that counter violence and thus attack more than they are attacked (e.g. militaries). The fifth category includes violent extremist groups that attack more than they are attacked, such as VPOs. Groups that are net attackers are indistinguishable at the level of individual behavior, but clearly separate into pro- and anti-violent extremism based on the groups to which they are close. This conclusion is in line with our original assumption that the propensity to use political violence concur with a group's position in the social network.

The second part of the article mapped a series of 389 events related to nine major Islamist groups in the region. Spatial analysis suggests that violent events involving Islamist groups have followed different patterns depending on the period under consideration but reveals no evidence of a border 'sanctuary'. While violence was concentrated almost exclusively within Algeria until 2004, cross-border movement has since intensified, following the establishment of military bases by AQIM in Mali. This suggests a 'mobility' scenario similar to the Arab revolt of the 20[th] century during which a highly mobile irregular force defeated the immobile and defensive Ottoman Turkish army. Our analysis suggests that until the French-led military offensive of 2013, military operations of trans-Saharan Islamist groups were "more like naval warfare than ordinary land operations, in their mobility, their ubiquity, their independence of bases and communications, their lack of ground features, of fixed directions, of fixed points" (Lawrence 1920). More recently, Islamist groups have concentrated their operations in Northern Mali as well as Southern Algeria, leaving Mauritania, Niger and Chad relatively unscathed. Owing to the Malian conflict and to a series of state and international military initiatives, cross-border movement has been on the wane in some countries, which seems to validate our original assumption that Islamist groups concentrate on border segments that are less heavily guarded and/or where informal arrangements with border officials are possible.

Our results have policy implications for governments and external forces involved in deterring politically violent organizations. First, unlike their adversaries, VPOs are socially and spatially connected across the regions; so, there is a need for collective security institutions that can help countries coordinate, build trust, and go beyond ad hoc engagements. In recent years, several 'Sahel' strategies have been initiated by organizations as diverse as the European Union (2011), the United Nations (2013), the Economic Community of West African States (2014), the African Union (2014), and the regional coordination framework G5 Sahel, to address governance, security and development in the region. Building institutional capacity around common interests is likely to pay off in a region that is largely devoid of collective security institutions. Precedent also suggests that states outside the region will continue to play a supporting rather than a lead role. In addition to supporting capacity-building efforts already underway, Western governments should be prepared to mount a comprehensive Whole-of-Government effort in support of local authorities that will minimize their local footprint while optimizing outcomes. From a military perspective, the fluidity of personal allegiances and mobility of actors across borders in the region calls for a mobile and flexible military response. Regional volatility notwithstanding, operations Serval and Barkhane suggest that desert insurgents are not impervious to external attack. As Western armies and their African allies become more mobile and flexible in their regional responses to political violence, desert insurgency proves to be a double-edged sword that can also work against those who know the terrain best.

**References**

ACLED (2015) Armed Conflict Location and Event Dataset, version 5, www.acleddata.com/data.

Arsenault, E.G. & Bacon, T. (2015) Disaggregating and defeating terrorist safe havens. *Studies in Conflict and Terrorism* 38: 85–112.

Author (2010)

Author (2013)

Author (2014)

Author (2015)

Arango, T. & Barnardmay, A. (2015) With victories, ISIS dispels hope of a swift decline. *The New York Times*, May 23.

Bøås, M. (2015) *The Politics of Conflict Economies*. London, Routledge.

Bøås, M. (2014) Guns, money and prayers: AQIM's blueprint for securing control of Northern Mali. *CTC Sentinel* 7(4): 1–5.

Brafman Kittner, C.C. (2007) The role of safe havens in Islamist terrorism. *Terrorism and Political Violence* 19: 307–329.

Burt, R.S. (1992) *Structural Holes. The Social Structure of Competition*. Cambridge, Harvard University Press.

Campana, A. & Ducol, B. (2011) Rethinking terrorist safe havens: Beyond a state-centric approach. *Civil Wars* 13(4): 396–413.

Csaba, L.Z. & J. Pál (2010) How negative networks are forming and changing in time? Theoretical overview and empirical analysis in two high-school classes. *Review of Sociology* 2: 70−96.

CSIS (2014) *Political Stability and Security in West and North Africa*. Ottawa, Cabinet Office, Canadian Security Intelligence Service.

De Juan, A. (2015) Long-term environmental change and geographical patterns of violence in Darfur, 2003–2005. *Political Geography* 45: 22−33

Doreian, P. & Krackhardt, D. (2001) Pre-transitive balance mechanisms for signed networks. *Journal of Mathematical Sociology* 25: 43−67.

Doreian, P. & Mrvar, A. (2015) Structural balance and signed international relations. *Journal of Social Structure* 16(2), https://www.cmu.edu/joss/content/articles/volume16/DoreianMrvar.pdf.

Dowd, C. (2015) Cultural and religious demography and violent Islamist groups in Africa. *Political Geography* 45: 11−21.

Dowd, C. (2013) Tracking Islamist militia and rebel groups. *Climate Change and African Political Stability*. http://strausscenter.org/ccaps/publications/researchbriefs.html.

Dowd, C, & Raleigh, C. (2013) Sahel state political violence in comparative perspective. *Stability* 2(2): 1−11.

Everett, M.G. & Borgatti, S.P. (2014) Networks containing negative ties. *Social Networks* 38: 111−120.

Flint, C., Diehl, P., Scheffran, J., Vasquez, J. & Chi, S. (2009) Conceptualizing ConflictSpace: Toward a geography of relational power and embeddedness in the analysis of interstate conflict. *Annals of the Association of American Geographers* 99(5): 827–835.

Forest, J.J. (2012) *Confronting the Terrorism of Boko Haram in Nigeria*. MacDill Air Force Base, The JSOU Press.

Freeman, L.C. (1979) Centrality in social networks: Conceptual clarification. *Social Networks* 1: 215‑239.

Galbert de, S. (2015) Does France have the firepower to fight the Islamic State? *Foreign Policy*, Nov 20, http://foreignpolicy.com/2015/11/20/france-military-fight-the-islamic-state-on-its-own-eu-us/

GCTAT (2013) *Organisation al-Qa'eda in the Lands of Islamic Maghreb*. Geneva, Geneva Center for Training and Analysis of Terrorism.

Global Terror Watch (2013) Base du Djihad au Maghreb Islamique (BDMI). http://www.globalterrorwatch.ch/?p=4411

Gow, J., Olonisakin, F., Dijxhoorn E (eds) *Militancy and Violence in West Africa*. London and New York, Routledge.

Grosser, T., Kidwell-Lopez, V. & Labianca, G. (2010) A social network analysis of positive and negative gossip in organizational life. *Group & Organization Management* 35: 177‑214.

Guidère, M. (2011) The tribal allegiance system within AQIM. *CTC Sentinel* 4(2): 9–11.

Hagen, A. (2014) *Al Qaeda in the Islamic Maghreb. Leaders and their Networks*. Washington D.C, American Enterprise Institute.

Herbst, J. (2002) *States and Power in Africa. Comparative Lessons in Authority and Control*. Princeton, Princeton University Press.

Huitsing, G., van Duijn, M.A.J., Snijders, T.A.B., Wang, P., Sainio, M., Salmivalli, C. & Veenstra, R. (2012) Univariate and multivariate models of positive and negative networks: Liking, disliking, and bully–victim relationships. *Social Networks* 34(4): 645–657.

Hummon, N.P. & Doreian, P. (2003) Some dynamics of social balance processes: bringing Heider back into balance theory. *Social Networks* 25(1): 17–49.

Innes, M.A. (ed.) (2007) *Denial of Sanctuary. Understanding Terrorist Safe Havens*. Westport, Praeger Security International.

Joscelyn, T. (2015) Mokhtar Belmokhtar now leads 'Al Qaeda in West Africa'. *The Long War Journal*, August 13, http://www.longwarjournal.org/archives/2015/08/mokhtar-belmokhtar-now-leads-al-qaeda-in-west-africa.php

Keegan, J. (1993) *A History of Warfare*. New York, Alfred Knopf.

Korteweg, R. & Ehrhardt, D. (2006) *Terrorist Black Holes: A Study into Terrorist Sanctuaries and Governmental Weakness*. The Hague, Clingendael Center for Strategic Studies.

Labianca, G. & Brass, D.J. (2006) Exploring the social ledger: Negative relationships and negative asymmetry in social networks in organizations. *Academy of Management Review* 31: 596‑614.

Lacher, W. (2012) *Organized Crime and Conflict in the Sahel-Sahara Region*. Washington D.C, Carnegie Endowment for International Peace.

Lawrence, T.E. (1920) The evolution of a revolt. A*rmy Quarterly and Defense Journal*, October.

Lebovich, A. (2013) The local face of jihadism in Northern Mali. *CTC Sentinel* 6(6): 4‑9.

van Mastrigt, S.B. & Carrington, P.J. (2014) Sex and age homophily in co-offending networks: opportunity or preference? In: Morselli, C. (ed.) *Crime and Networks*. Abingdon: Routledge, 28‑51.

Maoz, Z. (2006) Network polarization, network interdependence, and international conflict, 1816-2002. *Journal of Peace Research* 43(4): 391‑411.

Maoz, Z., Ranan, D., Kuperman, L., Terris, L. & Talmud, I. (2006) Structural equivalence and international conflict: a social network analysis. *Journal of Conflict Resolution* 50: 664–689.

Medina, R.M. & Hepner, G.F. (2013) *The Geography of International Terrorism. An Introduction to Spaces and Places of Violent Non-State Groups*. Boca Raton, CRC Press.

Menner, S. (2014) Boko Haram's regional cross-border activities. *CTC Sentinel* 7(10): 10–15.

Meservey, J. (2013) Al-Shabaab's Somali safe havens: A springboard for terror. *Perspectives on Terrorism* 7(6): 90–99.

Newman, M.E.J. (2010) *Networks. An Introduction*. Oxford, Oxford University Press.

OECD (2014) *An Atlas of the Sahara-Sahel: Geography, Economics and Security*. Paris, OECD.

O'Loughlin, J. & Raleigh, C. 2008. The spatial analysis of civil war violence. In: K. Cox, M. Low and J. Robinson (eds) *A Handbook of Political Geography*. Thousand Oaks, Sage: 493–508.

Ould Ahmed Salem, Z. (2013) *Prêcher dans le Désert: Islam Politique et Changement Social en Mauritanie*. Paris, Karthala.

Phillips, B.J. (2015) Enemies with benefits? Violent rivalries and terrorist group longevity. *Journal of Peace Research* 52(1): 62–75.

Phillips, M.D. & Kamen, E.A. 2014. Entering the black hole: The Taliban, terrorism, and organised crime. *Journal of Terrorism Research* 5(3), http://jtr.st-andrews.ac.uk/articles/10.15664/jtr.945/.

Radil, S.M., Flint, C. & Tita, G.E. (2010) Spatializing social networks: using social network analysis to investigate geographies of gang rivalry, territoriality, and violence in Los Angeles. *Annals of the Association of American Geographers* 100(2): 307–326.

Radil, S. & Flint, C. (2013) Exiles and arms: The territorial practices of state making and war diffusion in Post-Cold War Africa. *Territory, Politics, Governance* 1(2): 183–202.

Raleigh, C., Linke, A., Hegre, H. & Karlsen, J. (2010) Introducing ACLED: An armed conflict location and event dataset. *Journal of Peace Research* 47(5): 651–660.

Raleigh, C. & Dowd, C. (2015) Armed Conflict Location and Event Data Project (ACLED) Codebook, http://www.acleddata.com/wp-content/uploads/2015/01/ACLED_Codebook_2015.pdf

Rambaran, J.A., Dijkstra, J.K., Munniksma, A. & Cillessen, H.N. (2015) The development of adolescents' friendships and antipathies: A longitudinal multivariate network test of balance theory. *Social Networks* 43(4):162–176.

Smith, J.M., Halgin, D.S., Kidwell-Lopez, V., Labianca, G., Brass, D.J. & Borgatti, S.P. (2014) Power in politically charged networks. *Social Networks* 36: 162–176.

Tawil, C. (2009) *The Al-Qaeda Organisation in the Islamic Maghreb: Expansion in the Sahel and Challenges from Within Jihadist Circles?* Washington, DC, Jamestown Foundation.

The Punch (2012) We want Sharia law not independence – Malian Islamists. June 21, http://www.punchng.com/news/africa/we-want-sharia-law-not-independence-malian-islamists/

United Nations (2015) UN Security Council Committee pursuant to resolutions 1267 (1999) and 1989 (2011) concerning Al-Qaida and associated individuals and entities. New York, United Nations.

United States Department of State (2014) Country Reports on Terrorism 2013. Washington D.C., Bureau of Counterterrorism.

USAID (2014) *Assessment of the Risk of Violent Extremism in Niger.* Final Report. Washington D.C., United States Agency for International Development.

Uexkull, N. von (2014) Sustained drought, vulnerability and civil conflict in Sub-Saharan Africa. *Political Geography* 43: 16−26.

Wilkinson, H. (2012) Reversal of fortune: AQIM's stalemate in Algeria and its new front in the Sahel. In: SWAC/OECD (eds) *Global Security Risks and West Africa*. Paris, OECD: 11−33.

Wojtanik, A. (2015) *Mokhtar Belmokhtar: One-Eyed Firebrand of North Africa and the Sahel*. West Point, Combating Terrorism Center.

Zammit-Mangion, A., Dewar, M., Kadirkamanathan, V., Flesken, A. & Sanguinetti, G. (2013) *Modelling Conflict Dynamics with Spatiotemporal Data*. Berlin, Springer.

Zenn, J. (2012) *Northern Nigeria's Boko Haram: The Prize in al-Qaeda's Africa Strategy*. Washington D.C., The Jameston Foundation.

Zenn, J. (2014) Boko Haram: Recruitment, financing, and arms trafficking in the Lake Chad region. *CTC Sentinel* 7(10): 5−10.

Zenn J. (2015) Wilayat West Africa reboots for the Caliphate. *CTC Sentinel* 8(8): 10−16.

Zheng, Q., Skillicorn, D.B. & Walther, O. (2015) Signed directed social network analysis applied to group conflict. *IEEE International Conference on Data Mining*, DOI 10.1109/ICDMW.2015.107.

Zheng, Q. & Skillicorn, D.B. (2015) Spectral embedding of signed networks. *SIAM International Conference on Data Mining*: 55−63, DOI 10.1137/1.9781611974010.7.

Yap J. & Harrigan, N. (2015) Why does everybody hate me? Balance, status, and homophily: The triumvirate of signed tie formation. *Social Networks* 40: 103−122.

Appendix 1. Violent political organizations, 1997-2014

Abu Obeida Brigade
Abu Salim Martyrs' Brigade
Al Qaeda
Al Qaqa Brigade
Al-Burayqah Martyr's Brigade
Al-Salafiya
Al Jihadia
Ansar al-Sharia
Ansar Dine
Ansaru
AQIM: Al Qaeda in the Islamic Maghreb
Boko Haram
Brega Martyrs Brigade
El-Farouk Brigade
Falcons for the Liberation of Africa
February 17 Martyrs Brigade
Fighters of The Martyrs Brigade
FIS: Islamic Salvation Front
GIA: Armed Islamic Group
GMA: Mourabitounes Group of Azawad
GSL: Free Salafist Group
GSPC: Salafist Group for Call and Combat
Islamic Emirate of Barqa
Islamic State of Tripoli
Knights of Change
Libya Shield Brigade
LIDD: The Islamic League for Preaching and Holy Struggle
Martyrs' Brigade
MUJAO: Movement for Unity and Jihad in West Africa
Muslim Brotherhood
Nawasi Brigade
Nusur al-Sahel Brigade
Rafallah Sehati Brigade
Soldiers of the Caliphate in Algeria
Those Who Signed in Blood
Timizart Brigade

Appendix 2. Aggression levels for all groups that have more than one enemy

| Group | aggression | outaggression | inaggression |
|---|---|---|---|
| Rioters (Libya) | 1.32 | 1.32 | 0.00 |
| Military Forces of Libya | 0.43 | 0.50 | 0.07 |
| MUJAO | 0.32 | 0.84 | 0.53 |
| Military Forces of Algeria | 0.28 | 0.50 | 0.21 |
| Protesters (Mali) | 0.21 | 0.21 | 0.00 |
| Protesters (Libya) | 0.19 | 0.19 | 0.00 |
| GIA | 0.14 | 0.31 | 0.17 |
| Military Forces of Tunisia | 0.14 | 0.14 | 0.00 |
| Police Forces of Tunisia | 0.13 | 0.13 | 0.00 |
| Unidentified Armed Group (Algeria) | 0.11 | 0.11 | 0.00 |
| Civilians (France) | 0.11 | 0.26 | 0.15 |
| GMA | 0.10 | 0.16 | 0.06 |
| Libya Shield Brigade | 0.09 | 0.54 | 0.45 |
| Military Forces of Chad | 0.09 | 0.12 | 0.04 |
| Military Forces of Libya Special Forces | 0.06 | 0.13 | 0.06 |
| Libyan Rebel Forces | 0.06 | 0.27 | 0.20 |
| LIDD | 0.06 | 0.09 | 0.03 |
| GSPC | 0.05 | 0.31 | 0.26 |
| AQIM | 0.05 | 0.76 | 0.71 |
| Military Forces of Nigeria | 0.04 | 0.09 | 0.05 |
| Wershefana Communal Militia (Libya) | 0.03 | 0.14 | 0.10 |
| GLD | 0.01 | 0.04 | 0.04 |
| Boko Haram | 0.01 | 0.19 | 0.18 |
| Ansar Dine | 0.00 | 0.42 | 0.42 |
| El-Farouk Brigade | 0.00 | 0.06 | 0.06 |
| Military Forces of Mali | 0.00 | 0.22 | 0.22 |
| MNLA | 0.00 | 0.23 | 0.23 |
| Military Forces of France | 0.00 | 0.25 | 0.25 |
| Those Who Signed in Blood | -0.01 | 0.09 | 0.09 |
| Martyrs Brigade | -0.02 | 0.04 | 0.06 |
| February 17 Martyrs Brigade | -0.02 | 0.16 | 0.18 |
| Patriot Militia of Algerian Government | -0.03 | 0.12 | 0.15 |
| Abu Salim Martyrs Brigade | -0.04 | 0.00 | 0.04 |
| Civilians (Nigeria) | -0.04 | 0.05 | 0.09 |
| Military Forces of Mauritania | -0.05 | 0.05 | 0.10 |
| Military Forces of Niger | -0.07 | 0.10 | 0.17 |
| Al Qaqa Brigade | -0.07 | 0.00 | 0.07 |
| Ansaru | -0.08 | 0.04 | 0.12 |
| Police Forces of Algeria | -0.08 | 0.13 | 0.21 |
| FIS | -0.08 | 0.00 | 0.08 |
| Al Qaeda | -0.10 | 0.21 | 0.31 |
| Unidentified Armed Group (Libya) | -0.10 | 0.37 | 0.48 |
| Civilians (Libya) | -0.12 | 0.09 | 0.21 |
| Police Forces of Morocco | -0.12 | 0.00 | 0.12 |
| Civilians (Niger) | -0.12 | 0.00 | 0.12 |
| Civilians (Algeria) | -0.12 | 0.08 | 0.21 |
| Rafallah Sehati Brigade | -0.13 | 0.00 | 0.13 |
| UN | -0.14 | 0.07 | 0.21 |
| Zawia Ethnic Militia (Libya) | -0.16 | 0.00 | 0.16 |
| Civilians (Morocco) | -0.16 | 0.00 | 0.16 |
| Civilians (International) | -0.20 | 0.04 | 0.24 |
| Soldiers of the Caliphate in Algeria | -0.21 | 0.00 | 0.21 |
| Civilians (Mali) | -0.36 | 0.00 | 0.36 |
| Ansar al-Sharia | -0.52 | 0.53 | 1.05 |
| Muslim Brotherhood | -0.91 | 0.00 | 0.91 |

Note: For each group, the outaggression is the average length of outgoing attack ties in the embedded graph, inaggression is the average length of incoming attack ties, and net aggression is the difference of the two.

Appendix 3. Aggression levels for Libyan organizations

| Group | aggression | outaggression | inaggression |
|---|---|---|---|
| Protesters | 1.24 | 1.24 | 0.00 |
| Military Forces | 0.77 | 0.83 | 0.07 |
| Libyan Rebel Forces | 0.41 | 0.68 | 0.28 |
| Libya Shield Brigade | 0.36 | 1.13 | 0.77 |
| Wershefana Communal Militia | 0.20 | 0.51 | 0.31 |
| Misratah Communal Militia | 0.00 | 0.32 | 0.32 |
| Abu Salim Martyrs Brigade | 0.00 | 0.06 | 0.06 |
| Ansar al-Sharia | -0.04 | 0.70 | 0.74 |
| Islamist Militia | -0.06 | 0.00 | 0.06 |
| Brega Martyrs Brigade | -0.06 | 0.00 | 0.06 |
| Vigilante Militia | -0.07 | 0.00 | 0.07 |
| Al Qaeda | -0.09 | 0.00 | 0.09 |
| Al Qaqa Brigade | -0.11 | 0.85 | 0.96 |
| Zintan Ethnic Militia | -0.22 | 0.00 | 0.22 |
| Operation Libya Dawn | -0.23 | 0.00 | 0.23 |
| Zawia Ethnic Militia | -0.25 | 0.00 | 0.25 |
| Civilians | -0.26 | 0.22 | 0.48 |
| Gharyan Communal Militia | -0.30 | 0.00 | 0.30 |
| February 17 Martyrs Brigade | -0.47 | 0.06 | 0.53 |
| Rafallah Sehati Brigade | -0.81 | 0.00 | 0.81 |

Source: ACLED. Calculations: authors. Note: the following organizations with entirely zero rows have been removed from the table: El-Farouk Brigade, Al-Sawaiq Battalion, BSRC, Journalists, Police Forces, Salafist Group, Mutiny of Military Forces, Janzur Communal Militia, Awlad Suleiman Ethnic Militia, Shura Council of Benghazi Revolutionaries.

Appendix 4. Aggression levels for Nigerian organizations

| Group | aggression | outaggression | inaggression |
|---|---|---|---|
| Boko Haram | 1.08 | 3.51 | 2.43 |
| Martyrs Brigade | 0.98 | 0.98 | 0.00 |
| Ansaru | 0.72 | 1.09 | 0.36 |
| Shuwa Ethnic Militia | 0.48 | 0.48 | 0.00 |
| Lassa Communal Militia | 0.47 | 0.47 | 0.00 |
| Kawuri Communal Militia | 0.42 | 0.42 | 0.00 |
| Military Forces | 0.36 | 0.41 | 0.04 |
| Attagara Communal Militia | 0.32 | 0.32 | 0.00 |
| Civilian Joint Task Force | 0.04 | 0.04 | 0.00 |
| Borno Vigilance Youths Group | 0.04 | 0.04 | 0.00 |
| Unidentified Armed Group | 0.02 | 0.02 | 0.00 |
| Civilians (Lebanon) | 0.00 | 0.00 | 0.00 |
| Police Forces | 0.00 | 0.04 | 0.04 |
| Vigilante Militia | 0.00 | 0.04 | 0.04 |
| Militia (Ali Kwara) | 0.00 | 0.00 | 0.00 |
| Fulani Ethnic Group | 0.00 | 0.00 | 0.00 |
| All Progressives Congress | 0.00 | 0.00 | 0.00 |
| VGN: Vigilante Group | 0.00 | 0.48 | 0.48 |
| Shiite Muslim Group | 0.00 | 0.00 | 0.00 |
| Civilians (China) | -0.03 | 0.00 | 0.03 |
| UN | -0.04 | 0.00 | 0.04 |
| Christian Militia | -0.31 | 0.00 | 0.31 |
| Civilians (International) | -0.57 | 0.00 | 0.57 |
| Military Forces Joint Task Force | -0.64 | 0.00 | 0.64 |
| Civilians (South Korea) | -0.70 | 0.00 | 0.70 |
| Private Security Forces | -0.78 | 0.00 | 0.78 |
| Civilians (Europe) | -0.89 | 0.00 | 0.89 |
| Civilians | -0.98 | 0.04 | 1.02 |

Source: ACLED. Calculations: authors. Note: the following organizations with entirely zero rows have been removed from the table: ANPP, Prison Guards, Muslim Group, Christian Group, Students, PDP, Government, Military Forces UK, Igbo Ethnic Group.